\documentclass[12pt]{iopart}

\usepackage{iopams}  
\usepackage{graphicx}
\usepackage{adjustbox}
\usepackage[graphicx]{realboxes}
\usepackage{rotating}
\usepackage{epstopdf}
\usepackage{adjustbox}
\usepackage{float}
\usepackage{latexsym}
\usepackage{amssymb}
\usepackage{array}
\usepackage{longtable}

\begin{document}

\title[ Compact Objects in $f(R,T)$ gravity]{Anisotropic Compact Objects in Modified  $f(R,T)$ gravity}

\author{S  Dey$^1$, A Chanda$^2$, B C Paul$^3$}

\address{Department of Physics, University of North Bengal, Siliguri, Dist.  Darjeeling 734 013, West Bengal, India \\
      IUCAA Centre for Astronomy Research and Development, North Bengal  \\
E-mail : $^1$sagardey1992@gmail.com, $^2$anirbanchanda93@gmail.com, $^3$bcpaul@associates.iucaa.in}

\vspace{10pt}

\begin{abstract}
We obtain a class of anisotropic spherically symmetric relativistic solutions of compact objects in hydrostatic equilibrium  in the $f(R,T) =R+2\chi T$ modified gravity, where $R$ is the Ricci scalar, $T$ is the trace of the energy momentum tensor and $\chi$ is a dimensionless coupling parameter.  The matter Lagrangian is $L_{m}=- \frac{1}{3}(2p_{t}+p_{r})$, where $p_r$ and $p_t$ represents the radial and tangential pressures. 
Compact objects with dense nuclear matter is expected to be anisotropic.  Stellar models are constructed  for anisotropic neutron stars working in the modified Finch-Skea (FS) ansatz without preassuming an equation of state.  
 The stellar models are investigate plotting  physical quantities like energy density, anisotropy parameter, radial and tangential pressures in all particular cases. 
The stability of stellar models are checked  using  the causality conditions and adiabatic index. Using the observed mass of a compact star we obtain stellar models that predicts the radius of the star and  EoS for matter inside the  compact objects with different values of gravitational coupling constant $\chi$.  It is also found that  a more massive star can be accommodated with  $\chi <0$. The stellar models obtained here obey the physical acceptability criteria which show consistency for a  class of stable compact objects in modified $f(R, T)$ gravity. 
\end{abstract}

\maketitle

\section{Introduction}

 General theory of Relativity (GTR) is a geometric theory of gravitation formulated on the concept that gravity manifests itself as the curvature of space-time. Although GTR is a fairly successful theory at low energy,  it is entangled with some serious issues at ultraviolet and infrared limits. Some of the astronomical observational evidences namely,  Galactic, extra Galactic and cosmic dynamics are not understood in the framework of GTR. 
 The needle of hope points to the concept of  the existence of  exotic matter that represents the dark energy \cite{1,2} which we need if matter sector of GTR is to be modified.  On the other hand a  modification of  the gravitational sector to fit the missing matter-energy of the observed universe is also another important area of present research.
 In the  literature  \cite{3,4,5,6} a number of theories of gravity with modification of the gravitational sector came up  to understand the evolution of the observed universe as well as to solve some of the issues of non- renormalizability \cite{7,8}  in GTR.  In 1970, Buchdahl \cite{9} using a non-linear function of Ricci  scalar namely, $f(R)$ gravity theory  first introduced a modification of theory of gravity to explain some of the drawbacks in Friedmann- Lema\^{i}tre-Robertson-Walker cosmological models.  A higher derivative term in the gravitational action in the form $R^2$-term  was considered by Starobinsky \cite{10} and found  the existence of inflationary solutions in cosmology. \\
 Recently, Harko and his collaborators \cite{11}  introduced a more generalized form of gravity,  the $f(R)$ -gravity  which consists of a self-assertive expression of the Ricci scalar ($R$) and the trace of the energy-momentum tensor ($T$) together introducing $f(R,T)$-theory of  gravity. 
The modified theory is interesting as it is effectively  accommodate the late time acceleration of the universe. Consequently, there is a spurt in research activities in understanding  astrophysical objects of interest in the modified theory of gravity.  It is known that the presence of an extra force perpendicular to the four velocity in the $f(R,T)$ gravity helps test particles to follow a non-geodesic trajectory.  It is  shown \cite{12}  that for a specific linear form of  $f(R,T)$ -theory, say $f(R,T)= R + f(T)$, the trajectory of the particles become a geodesic path.  It is known \cite{13}  that that the $f(R,T)$ theory of gravity pass solar system test satisfactorily.
 A number of cosmological models \cite{17,18,19} are constructed in   the $f(R,T)$  theory of gravity which accommodates the observed universe successfully. Consequently,  Moraes et al. \cite{20} studied the equilibrium configuration of quark stars with MIT bag mode. 
 It is shown \cite{22} that an analytical stellar model for compact star in  $f(R,T)$ gravity may be obtained considering a correct form of the Tolman-Oppenheimer-Volkoff (TOV) equation. Deb {\it et al.} \cite{23} analyzed both isotropic and anisotropic spherically symmetric compact stars and presented the graphical analysis of LMC X-4 star model. 
The effect of higher curvature terms present in $f(R,T)$ gravity is probed in  compact objects \cite{24} making use of EoS given by polytropic and MIT bag model. 
 The physical properties of a star in the above case can be derived  knowing  EoS, $ i.e.$, $ p= p (\rho)$ which is not yet known for a compact object at extreme terrestrial condition. In the absence of a reliable information of the EoS at very high densities, assumption of the metric potentials, based on the geometry has been found to be a reasonable approach to construct a stellar model \cite{201,20a,20b,20c}.
The compact objects are stable objects at extreme terrestrial conditions.  Thus the compact object can be probed alternatively, where for a given geometry the EoS can be predicted. 
The motivation of the present paper is to obtain relativistic solution for  anisotropic compact stars with its  interior space-time described by Finch-Skea (FS) geometry in a linear modified $f(R,T)$ gravity and construct stellar models.
FS metric originated to  correct the Dourah and Ray \cite{25} metric which is not suitable for compact object, 
Finch and Skea \cite{26} modified the metric to describe  relativistic  stellar models. Subsequently, FS metric with a modification in  4- dimensions \cite{27,28,29} and  in higher dimensions \cite{30,31,32} are considered   to explore astrophysical objects. 
In compact objects the interior  pressure may not be same in all directions, thus the study of the behaviour of anisotropic pressure  for a spherically symmetric stellar model is important to explore.   Ruderman \cite{35} shown that at high density ($> 10^{15} g/cm^{3}$) nuclear matter object may be treated relativistically which exhibits the property of anisotropy. The reason for incorporating anisotropy is due to the fact that in the high density regime of compact stars the radial pressure ($p_{r}$) and the transverse pressure ($p_{t}$ ) are not equal which was pointed out by Canuto \cite{36}.  There are other reasons to assume anisotropy in compact stars which might occur in astrophysical objects for various reasons namely, viscosity, phase transition, pion condensation, the presence of strong electromagnetic field, the existence of a solid core or type 3A super fluid, the slow rotation of fluids etc. 
In this paper we construct relativistic stellar models and predict  EoS in the framework of 
 a linear $f(R,T)$ gravity  with  isotropic or anisotropic  fluid distribution.  \\
 
The outline of the paper is  as follows: in section $\bf 2$ we present the basic mathematical formulation of $f(R,T)$ theory and the field equations. In section $\bf 3$, a class of relativistic solutions are obtained  for different parameters of the theory. In section $\bf 4$ the constraints to obtain stellar models are presented. In section $\bf 5$, 
general properties of compact stars, the stability of stellar models, energy conditions, mass to radius etc. are discussed. 
 The EoS of mater inside the star is also predicted. Finally,  we discuss the results in section $\bf 6$.\\

\section{ The Gravitational action and the field equations in  $f(R,T)$ gravity}

The gravitational action for modified theory of gravity is given by
\begin{equation}
S=\frac{1}{16\pi} \int{f(R,T) \sqrt{-g} \; d^{4}x} +\int{\it{L}_{m}}\sqrt{-g} \; d^{4}x,
\end{equation}
where $f(R,T)$ is an arbitrary function of the Ricci scalar  $(R)$ and $(T)$ is the trace of the energy-momentum tensor $T_{\mu \nu}$. The determinant of the metric tensor $g_{\mu \nu}$ is given by $g$ and $L_{m}$ is the Lagrangian density of the matter part. We consider gravitation unit  $c=G=1$.
The field equations for the modified gravity theory can be obtained by varying the action $S$ with respect to the metric tensor $g_{\mu \nu}$ which is given by,
\[
(R_{\mu\nu}-\nabla_{\mu}\nabla_{\nu})f_{R}(R,T)+ g_{\mu \nu}\; \Box f_{R}(R,T) 
\]
\begin{equation}
\label{2}
-\frac{1}{2} g_{\mu \nu}\; f(R,T) =8\pi \; T_{\mu \nu}-f_{T}(R,T) \; (T_{\mu \nu}+\Theta_{\mu \nu}),
\end{equation}
where $f_{R}(R,T)$ denotes the partial derivative of $f(R,T)$ with respect to $R$, and $f_{T}(R,T)$ denotes the partial derivative of $f(R,T)$ with respect to $T$. $R_{\mu \nu}$ is the Ricci tensor, $\Box \equiv \frac{1}{\sqrt{-g}} \partial_{\mu}(\sqrt{-g}\; g^{\mu \nu} \partial_{\nu})$ is the D'Alembert operator and $\nabla_{\mu}$ represents the co-variant derivative, which is associated with the Levi-Civita connection of the metric tensor $g_{\mu \nu}$.
The energy momentum tensor $T_{\mu \nu}$ for perfect fluid changes the role in the $f(R, T)$-modified gravity  because of the presence of  $\nabla_{\mu} \nabla_{\nu} R$ and $(\nabla_{\mu}R)(\nabla_{\nu}R)$ and terms which originate  from trace of the energy momentum tensor $T$ in the field equation. In the paper, we consider compact objects with anisotropic matter distribution in the modified gravity.
 The stress-energy tensors $T_{\mu \nu}$ and $\Theta_{\mu \nu}$ are defined as,
\begin{equation}
T_{\mu \nu}=g_{\mu \nu} L_{m}-2 \;  \frac{\partial L_{m}}{\partial g^{\mu \nu}},
\end{equation}
\begin{equation}
\Theta_{\mu \nu}=g^{\alpha \beta} \; \frac{ \delta T_{\alpha \beta}}{\delta g^{\mu \nu}}.
\end{equation}
Using  eq. (2) the covariant divergence of the stress-energy tensor can be written as
\begin{equation}
\nabla^{\mu}T_{\mu \nu}=\frac{f_{T}}{8\pi-f_{T}}  \left[( T_{\mu \nu}+\Theta_{\mu \nu})\nabla^{\mu}ln f_{T}
+\nabla^{\mu}\Theta_{\mu \nu}-\frac{1}{2} \; g_{\mu \nu}\nabla^{\mu}T \right].
\end{equation}
It may be mentioned here that the covariant derivative of the stress-energy tensor in $f(R,T)$ theory  does not vanishes, which is different from the $f(R)$-theory. Consequently we describe an effective energy density and pressure which however leads $T^{eff}_{\mu \nu};\mu=0$.

In the modified gravity  $f(R,T)=R+2 \, \chi T$, where $\chi$ is a coupling constant,  the field eq. (\ref{2}) can be represented as 
\begin{equation}
G_{\mu \nu}= 8\pi \; T^{eff}_{\mu\nu}
\end{equation}
where $G_{\mu \nu}$ is the  Einstein tensor and $T^{eff}_{\mu\nu}$ is the effective energy-momentum tensor.
The energy-momentum tensor for  anisotropic matter distribution is given by 
\begin{equation}
T_{\mu \nu}=(\rho + p_{t}) \; u_{\mu}u_{\nu}-p_{t} \; g_{\mu \nu}+(p_{r}-p_{t}) \; v_{\mu}v_{\nu}
\end{equation}
where $v_{\mu}$ is the radial four-vector, while $u_{\nu}$ is four velocity vector,  $\rho$, $p_{r}$ and $p_{t}$ are  the  energy density, the radial and tangential pressures respectively.  Here, we consider the matter Lagrangian $L_{m}=- P$, where $P=\frac{1}{3}(2p_{t}+p_{r})$.  For anisotropic fluid $\Theta_{\mu \nu}=-2T_{\mu \nu}-  g_{\mu \nu} P$,   the effective energy-momentum tensor becomes 
\begin{equation}
 T^{eff}_{\mu\nu}= T_{\mu \nu} \left(1+ \frac{\chi}{4 \pi} \right) + g_{\mu \nu}\; \frac{\chi}{ 8 \pi} \left( T + 2 P \right) .
\end{equation}
The above expression  contains the original matter stress-energy tensor $T_{\mu\nu}$ and the curvature terms \cite{11}. We consider  $f(R,T)=R+2 \, \chi T$ and  the eq.(5) becomes
\begin{equation}
\nabla^{\mu}T_{\mu \nu}= - \; \frac{\chi }{2( 4\pi+ \chi)} ( g_{\mu \nu} \nabla^{\mu}T + 2 \nabla^{\mu}(g_{\mu \nu} P) ).
\end{equation}
Now the effective conservation of energy equation is given by
\begin{equation}
\nabla^{\mu}\; T^{eff}_{\mu\nu} = 0.
\end{equation}
Thus the modified gravity allows a non-linear regime in addition to linear regime effectively.   The motivation of the paper is to  study the characteristics of gravitational dynamics in the compact objects having density greater than the nuclear density  in the $ f(R,T)$- theory of gravity which is an extension of both GTR and  $f(R)$-gravity. 

\section{Modified Field Equations in $f(R,T)$ gravity} 
We consider a  spherically symmetric metric for the interior spacetime of a static stellar configuration given by 
\begin{equation}
ds^{2}=e^{2\nu(r)}dt^{2}-e^{2\lambda(r)}dr^{2}-r^{2}(d\theta^{2}+sin^{2}\theta d\phi^{2}),
\end{equation}
where $\nu$ and $\lambda$ are the metric potentials which are functions of radial coordinate ($r$) only. The non-zero components of the energy momentum tensors are given by
\begin{equation}
T^{0}_{0}=\rho (r),
\end{equation}
\begin{equation}
T^{1}_{1}=-p_{r}(r),
\end{equation}
\begin{equation}
T^{2}_{2}=T^{3}_{3}=-p_{t}(r)
\end{equation}
where $p_r$ and $p_t$ are radial and tangential pressures respectively.  Using eqs. (6) - (8),  the field equations  can be rewritten as
\begin{equation}
e^{-2\lambda}\Big(\frac{2\nu'}{r}+\frac{1}{r^{2}}\Big)-\frac{1}{r^{2}}=  8\pi p^{eff}_{r},
\end{equation}
\begin{equation}
e^{-2\lambda}\Big(\nu''+\nu^{'2}+\frac{\nu'-\lambda'}{r}- \nu' \lambda'\Big) = 8\pi p^{eff}_{t},
\end{equation}
\begin{equation}
e^{-2\lambda}\Big(\frac{2\lambda'}{r}- \frac{1}{r^{2}}\Big)+\frac{1}{r^{2}}= 8\pi \rho^{eff},
\end{equation}
where the prime $(')$ is differentiation  w.r.t.  radial coordinate, $\rho^{eff}$, $p^{eff}_{r}$ and $p^{eff}_{t}$ are the effective density, radial pressure and tangential pressure. We get
\begin{equation}
  \rho^{eff} =  \rho + \frac{\chi}{24 \pi}(9 \rho - p_{r} - 2 p_{t})
\end{equation}
\begin{equation}
p^{eff}_{r} = p_{r}- \frac{\chi}{24 \pi}(3\rho - 7p_{r} - 2p_{t})
\end{equation}
\begin{equation}
p^{eff}_{t} = p_{t}- \frac{\chi}{24 \pi}(3\rho - p_{r} - 8 p_{t}).
\end{equation}
To study the matter content inside the compact objects, the field eqs.(15)-(17) are used to  determine the components of $T_{\mu\nu}$ $i.e.$ $\rho$, $p_{r}$ and $p_{t}$.
Using eqs.(15) and (16) we get a second order differential equation which is
\begin{equation}
 \nu ^{\prime \prime }+{\nu ^{\prime }}^{2}-\nu ^{\prime}\lambda ^{\prime }-\frac{\lambda ^{\prime }}{r}-\frac{\nu^{\prime }}{r}-\frac{1}{r^{2}}+\frac{e^{2\lambda}}{r^{2}}=2 (4 \pi +\chi) \; \Delta \;  e^{2\lambda}
\end{equation}
where  $\Delta= p_{t}- p_{r} $, which  represents the measure of anisotropy in pressure. In terms of effective pressures we get
\[
p^{eff}_{r} - p^{eff}_{t} = \left(1+ \frac{\chi}{4 \pi} \right) \; \Delta
\]
which is related to the anisotropy measure. It is evident that for isotropic pressure {\it i.e.}  $p_{r} = p_{t} $ one finds isotropy in the effective pressure. It  is also noted that the effective pressure difference vanishes  even if $p_{r} \neq p_{t} $ when $\chi = - 4 \pi$.
\\

In this section we adopt  following transformations first proposed by Durgapal and Bannerji \cite{db} on the matric potentials  to obtain relativistic solutions 
\[
A^{2}y^{2}(x)= e^{2\nu(r)}, \; \; \; \; \;  Z(x)=e^{2\lambda(r)}, \; \; \; \;  x= Cr^{2}.
\]
where $A$ and $C$ are arbitrary constants. The  above  transformation  reduces eq. (21)  to a second order differential equation which is given by  
\begin{equation}
4 x^{2}z \; \ddot{y}+2 x^{2}\; \dot{z}\, \dot{y}+y \left[x \, \dot{z}-z+1-\frac{2(4\pi +\chi) }{C} \,x \,\Delta\right]=0
\end{equation}
where the overdot denotes differentiation $w.r.t.$  the variable $x$.

\subsection{Exact Relativistic Solutions}

The  eq. (22) is further simplified introducing  $Z(x)$ \cite{tm} as
\begin{equation}
Z=\frac{1}{1+x}.
\end{equation}
Note that the choice of $Z$ is a sufficient condition for a static perfect fluid sphere  which is regular at the center  \cite{mm}.  Eq.(22) can be expressed as
\begin{equation}
4(1+x)\; \ddot{y}-2\; \dot{y}+(1-\alpha)\; y=0
\end{equation}
where
$ \alpha= \frac{ 2 \Delta (x+1)^2 (4 \pi + \chi)}{Cx} $. The measure of anisotropy is given by
\begin{equation}
\Delta = \frac{ \alpha \; x \; C}{2 (4 \pi + \chi) (x+1)^2}
\end{equation}
for $\chi \neq -4 \pi$ and $C\neq0$.   For $\alpha =0$, one recovers Finch-Skea model with an isotropic pressure distribution.  For anisotropic star, $\Delta$  vanishes  at the center   ($i.e.\;  p_{r}= p_{t}$), but away from the centre it is a regular solution which   grows showing different patterns of evolution for both the pressures. For $ -1 < \alpha < 1$,  we substitute the following :  $X= 1+ x$ and $y(x) = Z$  for simplicity in eq.(24) which yields 
\begin{equation}
\label{26}
4X\; \frac{d^{2}Z}{dX^{2}}- 2\; \frac{dZ}{dX}+(1-\alpha)\; Z=0.
\end{equation}
Once again we introduce the following  transformations: $Z= w (X) X^{n}$ and $u = X^{\gamma}$, where $\gamma $ and $n$ are real numbers. The above differential equation can be  reduced to a  standard Bessel equation.  
For $\gamma = \frac{1}{2}$ and $ n= \frac{3}{4} $,  the eq. (26) reduces to
\begin{equation}
\label{27}
u^{2}\frac{d^{2}w}{du^{2}}+ u \frac{dw}{du}+ \left[(1-\alpha) \; u^{2} - \frac{3}{2} \right]w=0.
\end{equation}
Now we consider further transformation from $u$ to $v$ variable as $(1-\alpha)^{\frac{1}{2}} u = v$  in the eq.(\ref{27})  which leads to a second order differential equation as follows
\begin{equation}
v^{2}\; \frac{d^{2}w}{dv^{2}}+ v \; \frac{dw}{dv}+\left[v^{2} -\left(\frac{3}{2} \right)^{2}\right]\;w=0.
\end{equation}
which is the Bessel equation of the order $\frac{3}{2}$. The general solution is given by 
\[
w= c_{1} \;  J_{\frac{3}{2}}(v) +  c_{2} \; J_{-\frac{3}{2}}(v) 
\]
where $ c_{1}$  and $ c_{2}$ are integration constants, $ J_{\frac{3}{2}}(v)$ and  $J_{-\frac{3}{2}}(v)$ are the Bessel functions, which can be written in terms of trigonometric functions. The general solution of the eq.(24) for  modified FS-metric in four dimension \cite{mm}   is given by
\[
y (x)= (1- \alpha)^{\frac{-3}{4}} [(b - a \sqrt{(1+C\;r^{2})(1- \alpha)}) 
\]
\[
\cos\sqrt{(1+C\;r^{2})(1- \alpha)}+ (a + b \sqrt{(1+C\;r^{2})(1- \alpha)}) 
\]
 \begin{equation}
   \sin\sqrt{(1+C\;r^{2})(1- \alpha)}]
\end{equation}
where, $a = c_{1} \sqrt{\frac{2}{\pi}} $ and $b = - c_{2} \sqrt{\frac{2}{\pi}} $ are arbitrary constants of the metric. 
We consider  the metric potential of the modified 4-dimensional FS -metric as
\begin{equation}
\label{28}
e^{2\lambda (r)}=  1+  C\; r^{2},
\end{equation}
\[
e^{2 \nu (r)}= (1- \alpha)^{\frac{-3}{2}} A^{2} [(b- a \sqrt{(1+C\;r^{2})(1- \alpha)}) 
\]
\[
 \cos \sqrt{(1+C\;r^{2})(1- \alpha)}+ (a+ b \sqrt{(1+C\;r^{2})(1- \alpha)}) 
 \]
\begin{equation}
 \label{29}
 \hspace{1.9 cm}  \; \sin\sqrt{(1+C\;r^{2})(1- \alpha)}]^{2}
\end{equation}
where $C$, $a$, $b$, $A$ and  $\alpha$  are the five unknowns. For $\alpha = 0$,  the Finch- Skea solution obtained in  GR for $4$ - dimensions  with isotropic fluid is recovered \cite{26}.
 The  relativistic solution  for $-1 < \alpha <1$ obtained here  is regular in the interior of the star which can be  matched smoothly with the Schwarzschild exterior solution at the boundary. It  can be used to construct stellar models determining the metric parameters $a$ , $b$ and $C$ for  given values of $\alpha$ and $\chi$.  It may be mentioned here that for $\alpha\geq 1$, the stellar models are not stable.
  Consequently, we consider  $-1<\alpha< 1$ in the $f(R,T)$- modified gravity to construct stellar models for compact objects.
 
 \section{ Analysis  for  Stellar Models}
 
The  following conditions \cite{gac} are imposed on the relativistic solutions for a physically realistic stellar configurations for compact objects in the modified gravity  :\\
$\bullet$ At the boundary of a static  star ($i.e.$ at $ r=b$), the interior space-time is matched with the exterior Schwarzschild solution. For the continuity of the metric functions at the surface, one consider
\begin{equation}
e^{2\nu (r)}|_{r=b}=\Big(1-\frac{2M}{b}\Big)
\end{equation}
\begin{equation}
e^{2\lambda (r)}|_{r=b}=\Big(1-\frac{2M}{b}\Big)^{-1}
\end{equation}
$\bullet$ The radial pressure ({\it $p_r$}) drops from its maximum value (at the center) to vanishing value at the boundary , $i.e.$, at $r= b$ , 
 $ p_{(r=b)}=0$, the radius of the star $b$ can be estimated.
\\
$\bullet$  The causality condition  is satisfied when  the speed of sound $ v^{2} = \frac{dp} {d\rho} \leqslant 1$ which is also a condition for stable stellar configuration \cite{33}.
\\
$\bullet$  The gradient of the pressure and energy-density should be negative inside the stellar configuration, $\it  i.e.$,  $\frac {dp_{r}}{dr}<  0$ and  $\frac {d\rho} {dr} < 0 $.
\\
$\bullet$  At the center of the star, $\Delta(0) = 0$ which implies zero radial and tangential pressure, $p_{r}(0) = p_{t}(0)$.
\\
$\bullet$  The anisotropic fluid sphere must satisfy the following three energy conditions, viz., (a) null energy condition (NEC), (b) weak energy condition (WEC) and (c) strong energy condition (SEC) if it is made up of normal fluid.
\\
$\bullet$  The adiabatic index : $ \Gamma = \frac{\rho+p}{p} \; \frac{dp}{d\rho} >  \frac{4}{3} $  required  for ensuring stability of the stellar configuration   \cite{34}. 

There are three field equations and five unknowns, to solve the equations  two $ad hoc$ assumptions are necessary for obtaining exact solutions.
Thus to  construct stellar models, the  unknown metric parameters $a$, $b$, $C$ for a given mass $(M)$ and radius $(r= b)$ of a star are to be determined from the boundary conditions making use of permissible values of $\alpha$ and $\chi$ for a realistic stellar model. Alternatively, for a given mass we can predict the radius of the compact objects for values of the other parameters.

\section{Physical Properties of compact stars for $-1  < \alpha <1$}

The physical features of anisotropic compact objects are  studied for $-1  < \alpha <1$. As the relativistic solutions are highly complex we analyze numerically the variations of the energy density, radial pressure, transverse pressures, energy conditions, anisotropy of pressure and stability for a given value of the model parameters. The graphical plots are important for predicting  the EoS of the observed compact objects. We consider uncharged anisotropic stellar objects. 
 
 \subsection{\bf Density and Pressure of a compact objects in $f(R,T)$ gravity}
 
 In the  $ f(R, T )$ - modified gravity we  determine physical parameters, namely,  energy density ($\rho$), radial pressure ($p_{r}$) and  tangential pressure ($p_{t}$).  The   metric potentials  $e^{2\lambda(r)}$   and $e^{2\nu(r)}$ given by eqs. (\ref{28}) and (\ref{29}) are employed in  eqs. (15) - (17) to determine the  energy density ($\rho$), radial pressure ($p_{r}$) and  tangential pressure ($p_{t}$) which are given by
\begin{equation}
\label{30}
\resizebox{0.5\textwidth}{!}{$   
\rho= \frac{C \left(\sin \left(\sqrt{{\bf h_1}}\right) \left(a \; {\bf h_2} +b h_o \sqrt{{\bf h_1}}   \right)+\cos \left(\sqrt{{\bf h_1}}\right) \left(b {\bf h_2} -a {\bf h_o}  \sqrt{\bf{ h_1}}  \right) \right)}{g(r,a,b, C, \chi) }
$},
\end{equation}
\begin{equation}
\label{31}
p_{r}= \frac{C \left( {\bf h_3} \cos \sqrt{{\bf h_1}} - {\bf h_4}  \sin \sqrt{{\bf h_1}} \right)}{g(r,a,b, C, \chi) },
\end{equation}
\begin{equation}
\label{32}
p_{t}= \frac{C \left(  {\bf h_5 } \sin \sqrt{{\bf h_1}}+ {\bf h_6 } \cos \sqrt{{\bf h_1}} \right)}{g(r,a,b, C, \chi) }.
\end{equation}
where the denominator is denoted as $g(r,a,b, C, \chi) = 
12 \left(\chi ^2+6 \pi \chi +8 \pi ^2\right) \left(C r^2+1\right)^2 (\sin \sqrt{{\bf h_1}} \left(a+b \sqrt{{\bf h_1}}\right)+\cos \sqrt{{\bf h_1}} \left(b-a \sqrt{{\bf h_1}}\right) )$,

${\bf h_o}=\chi  \left(C r^2 (\alpha +3)+12\right)+12 \pi  \left(C r^2+3\right)$,
${\bf h_1} = - (\alpha -1) (C r^2+1)$, 
${\bf h_2} = \chi  (-2 C r^2 (\alpha -3)-3 (\alpha -5))+12 \pi  \left(C  r^2+3\right)$,  
${\bf h_3} =a \sqrt{h_1}  \; (C r^2 (\alpha +3) \chi +12 \pi  (C r^2+1)) +b  \; (\chi  (2 C r^2 (3-5 \alpha )-9 \alpha +9)- 12 \pi  (2 \alpha -1) \;  C r^2+1))$,  
${\bf h_4}= a (\chi  (2 C r^2 (5 \alpha -3)+9 (\alpha -1))+12 \pi  (2 \alpha -1) (C r^2+1))+b \sqrt{h_1} (C r^2 (\alpha +3) \chi +12 \pi \;  (C r^2+1))$,  
    ${\bf h_5}= a (\chi  (2 C r^2 (3-2 \alpha )-9 \alpha +9)-12 \pi  (C r^2 (\alpha -1)+2 \alpha -1))+b \sqrt{h_1} (C r^2 (5 \alpha -3) \;  \chi +12 \pi  (C r^2 (\alpha -1)-1))$, 
   ${\bf h_6}= a \sqrt{h_1} (C r^2 (3-5 \alpha ) \chi -12 \pi  (C r^2 (\alpha -1)-1))+b (\chi  (2 C r^2 (3-2 \alpha )-9 \alpha +9)-12 \pi  (C r^2 (\alpha -1)+2 \alpha -1))$.
   \\

\begin{figure}
\centering
\includegraphics[width=8 cm,height= 5cm]{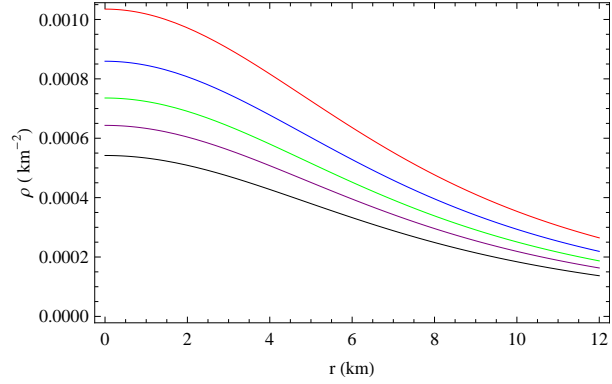}
\caption{Radial variation of energy-density ($\rho$) in PSR J0348+0432 for $\chi = 1$ (Red), $\chi= 3$ (Blue), $\chi = 5$ (Green), $\chi=7$ (Purple) and $\chi=10 $ (Black) (considering $\alpha=0.5$)}
\end{figure}

\begin{figure}
\centering
\includegraphics[width=8 cm,height= 5cm]{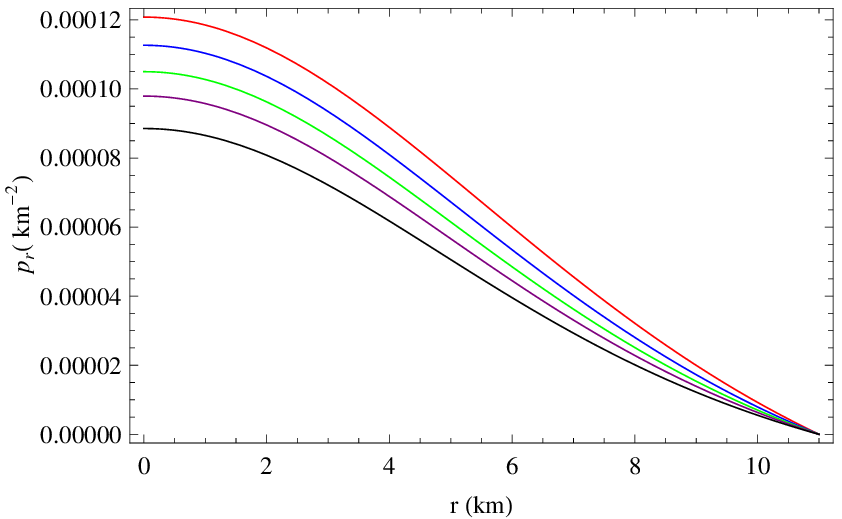}
\caption{Radial variation of radial pressure ($p_{r}$) in PSR J0348+0432 for $\chi = 1$ (Red), $\chi= 3$ (Blue), $\chi = 5$ (Green), $\chi=7$ (Purple) and $\chi=10 $ (Black) for $\alpha=0.5$}
\end{figure}
\begin{figure}
\centering
\includegraphics[width=8 cm,height= 5 cm]{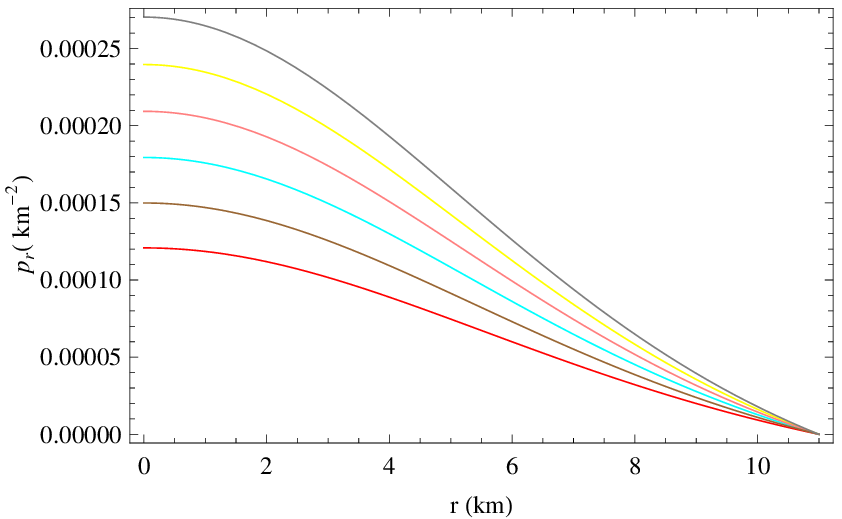}
\caption{Radial variation of radial pressure ($p_{r}$) in PSR J0348+0432 for $\alpha= 0$ (Gray), $\alpha= 0.1$ (Yellow), $\alpha = 0.2 $ (Pink), $\alpha=0.3$  (Cyan), $\alpha=0.4$(Brown) and $\alpha= 0.5 $ (Red) for $\chi=1$}
\end{figure}
\begin{figure}
\centering
\includegraphics[width=8 cm,height= 5 cm]{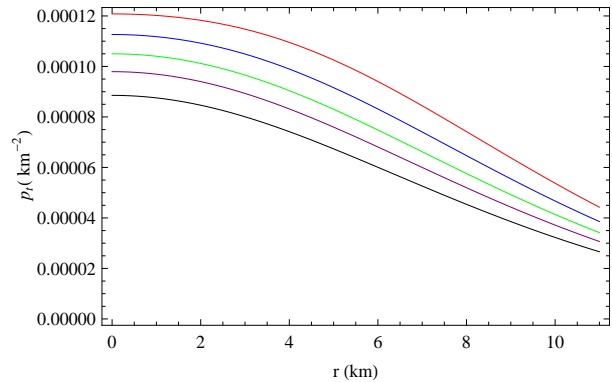}
\caption{Radial variation of transverse pressure ($p_{t}$) in PSR J0348+0432 for $\chi = 1$ (Red), $\chi= 3$ (Blue), $\chi = 5$ (Green), $\chi=7$ (Purple) and $\chi=10 $ (Black) for $\alpha=0.5$}
\end{figure}

The radial variation  of the energy density ($\rho$), radial pressure ($p_{r}$) and tangential pressure ($p_{t}$) are  plotted for  PSR J0348+0432 in Figs. (1),  (2) and (4) respectively for  $\alpha  =0.5$ with different  $\chi$.  It is evident that   the physical quantities are maximum at the origin which however, decrease monotonically away from the centre. As $\chi$ is increased the values of the physical parameters decreases.  Similarly,  the radial variation of radial pressure ($p_{r}$) and tangential pressure ($p_{t}$) for different $\alpha$  for  $\chi= 1$  are plotted  in Figs. (3) and (5) respectively. It is noted that as  $\alpha$ increases,  the  radial pressures and tangential pressure decreases which  are positive and regular at the origin with maximum values.  Thus the model is free from physical and mathematical singularities. It is also evident that the radial variation of energy density gradient and radial pressure gradient for different values of $\chi$ are plotted in Figs. (6) and (7) respectively,  which are found negative and it increases as $\chi$ is decreased for $\alpha =0.5$ . 
 \subsection{\bf Anisotropic Star}
 
Th anisotropy of a compact star which is determined by the difference of tangential and radial pressures is obtained from eqs. (\ref{31}) and (\ref{32}) as follows:
\begin{equation}
\label{33}
\Delta= p_{t} - p_{r} = \frac{C^2 r^2 \alpha }{2 (\chi +4 \pi ) \left(C r^2+1\right)^2}.
\end{equation}
An isotropic stellar model can be obtained for   $\alpha=0$  in $4$-dimensions, it is  evident that in modified gravity it always permits anisotropic star unless $\chi = - 4 \pi$ which follows from eq. (\ref{33}). 
In GTR, it is known that FS metric does not permit  anisotropic compact star in  a 4-dimensional geometry, but recently it is shown that a  higher dimensional extension of the Finch-Skea geometry permits an anisotropic star \cite{31}.   As the structure of $f(R,T)$ -gravity is interesting  found that anisotropic star is always permitted in a 4-dimensional FS metric.  In Fig. (8),  we plot the radial variation of  $\Delta$ for different $\chi$ values with a given $\alpha$.  It is found that for $\chi> 0$, $\Delta$ is positive {\it i.e.}, $p_{t}> p_{r}$ which in turn implies that the anisotropic stress is directed outwards, hence there exists a repulsive gravitational force that allows the formation of super massive stars.

The radial variation of $\Delta$ for different  values of $\alpha$ in the range $(- 1.0 \; to + \; 0.5)$ is drawn in Fig. (9) with $\chi=1$. It is evident that when $\chi =0$ it corresponds to isotropic star ($\alpha =0$) in four dimensions but when $\alpha <0$ one gets $\Delta <0$ for positive values of $\chi$ with $|\chi| \neq 4 \pi$.The anisotropy increases as    $\alpha$ increases but it decreases if $\alpha$ is more negative. It is also noted that for  negative values of $\alpha$ one gets $\Delta< 0$, where the radial pressure is greater than the tangential one, {\it i.e.}, $p_{r}> p_{t}$.
We  also note  that when $\alpha= - 0.5$,  we get a situation where $\Delta < 0 $ in the range  $-1> \chi> -3.5$  for PSR J0348 + 0432. Similarly,  $\Delta < 0 $ is recorded for $\alpha= - 0.2$, in the range $-1> \chi> - 4.4$ and for  $\alpha= - 1.0$, in the range  $-1> \chi> - 1.4$. Thus it is clear that this negative range of $\chi$ varies with negative $\alpha$ values which permits compact objects with $p_t > p_r$. However, our model is not allowed for negative  $\chi$ values with positive $\alpha$. 
\begin{figure}
\centering
\includegraphics[width=8 cm,height= 5 cm]{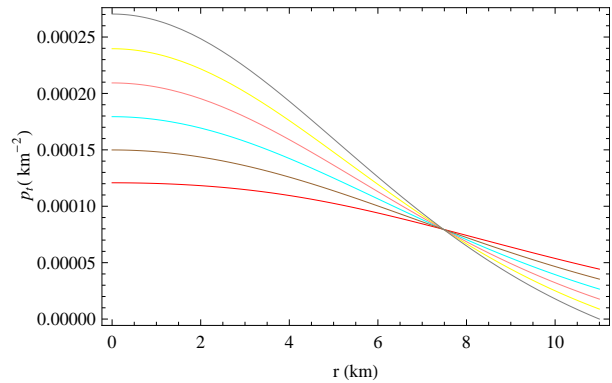}
\caption{Radial variation of transverse pressure ($p_{t}$) in PSR J0348+0432 for $\alpha= 0$ (Gray), $\alpha= 0.1$ (Yellow), $\alpha = 0.2 $ (Pink), $\alpha=0.3$ (Cyan), $\alpha=0.4$ (Brown) and $\alpha= 0.5 $ (Red) for $\chi=1$}
\end{figure}

\begin{figure}
\centering
\includegraphics[width=8 cm,height= 5 cm]{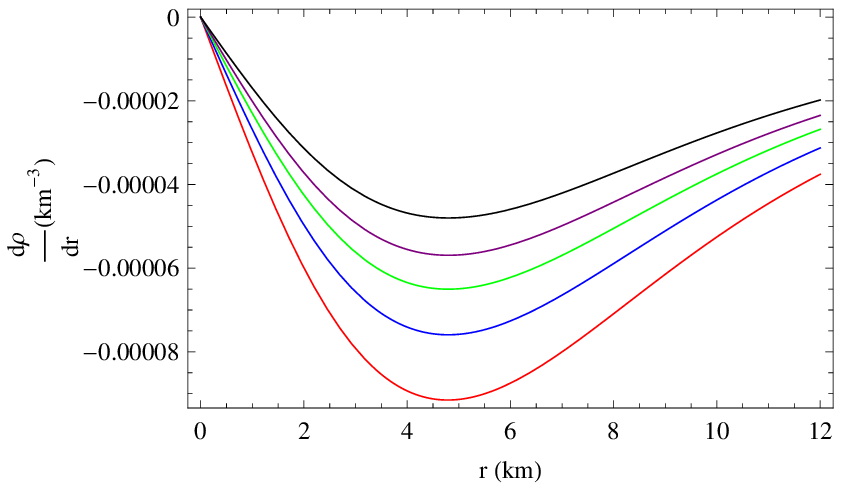}
\caption{Radial variation of energy-density gradient ($\frac{d\rho}{dr}$) in PSR J0348+0432 for $\chi = 1$ (Red), $\chi= 3$ (Blue), $\chi = 5$ (Green), $\chi=7$ (Purple) and $\chi=10 $ (Black)  for  $\alpha=0.5$}
\end{figure}

\begin{figure}
\centering
\includegraphics[width=8 cm,height= 5 cm]{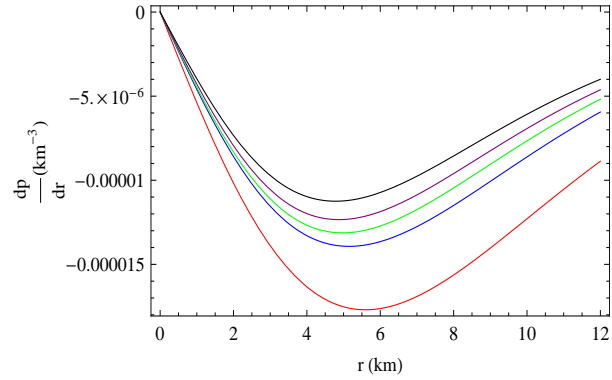}
\caption{Radial variation of pressure gradient ($\frac{dp_{r}}{dr}$) in PSR J0348+0432 for $\chi = 1$ (Red), $\chi= 3$ (Blue), $\chi = 5$ (Green), $\chi=7$ (Purple) and $\chi=10 $ (Black) for  $\alpha=0.5$}
\end{figure}

\begin{figure}
\centering
\includegraphics[width=8 cm,height= 5 cm]{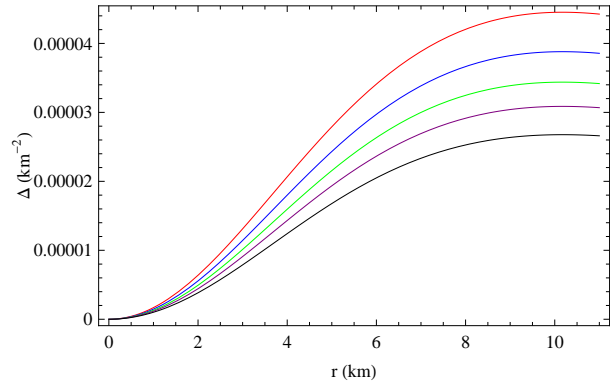}
\caption{Radial variation of anisotropy parameter ($\Delta$) in PSR J0348+0432 for $\chi = 1$ (Red), $\chi= 3$ (Blue), $\chi = 5$ (Green), $\chi=7$ (Purple) and $\chi=10 $ (Black) for  $\alpha=0.5$}
\end{figure}

\begin{figure}
\centering
\includegraphics[width=8 cm,height= 5 cm]{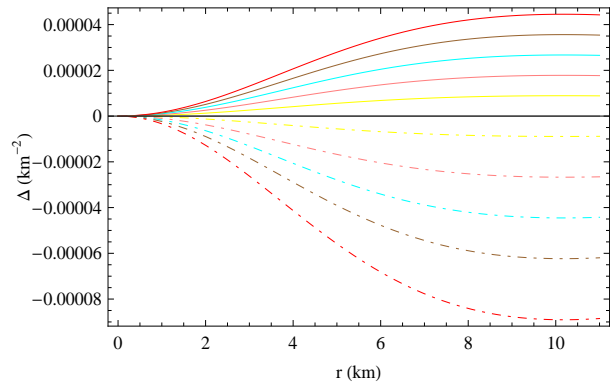}
\caption{Radial variation of anisotropy parameter ($\Delta$)  in PSR J0348+0432 for $\alpha= - 1.0$ (Red, DotDashed), $\alpha= - 0.7$ (Brown, DotDashed), $\alpha= - 0.5 $ (Cyan, DotDashed), $\alpha= - 0.3$ (Pink , DotDashed), $\alpha= - 0.1$ (Yellow,DotDashed), $\alpha= 0$ (Gray), $\alpha= 0.1$ (Yellow), $\alpha = 0.2 $ (Pink), $\alpha=0.3$  (Cyan), $\alpha=0.4$ (Brown) and $\alpha= 0.5 $ (Red) for $\chi=1$}
\end{figure}

\subsection{\bf Stability of the Stellar Model}
\subsubsection{\bf Herrera cracking concept}

The stability of a stellar model is studied numerically plotting the  radial variation of  the square of the radial speed of sound ($v_{r}^{2}= \frac{dp_{r}}{d\rho}$) and square of the transverse speed of sound ($v_{t}^{2}= \frac{dp_{t}}{d\rho}$) separately   in Figs. (10) and (11) respectively.  It is found that a stable configuration of anisotropic compact object can be accommodated. Herrera  and Abreu \cite{38}  pointed out that for a physically stable stellar system made of anisotropic fluid distribution the difference of square of the sound speeds should maintain its sign inside the stellar system.  Accordingly, in a  potentially stable region, square of the radial sound speed should be greater than the square of the tangential sound speeds. Hence, according to Herreras cracking conjecture  the required condition $|v_{t}^{2}- v_{r}^{2}| \leq 1$ is found to satisfy. 
We  plot variation of $|v_{t}^{2}- v_{r}^{2}|$ w.r.t. $r$ in Fig. (12) and it is found that  the condition is found to satisfy  $|v_{t}^{2}- v_{r}^{2}| \leq 1$ for different values of $\chi$ with $\alpha=0.5$.

\begin{figure}
\centering
\includegraphics[width=8 cm,height= 5 cm]{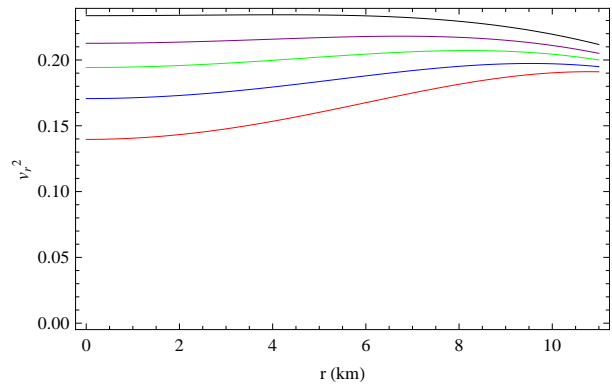}
\caption{Radial variation of $v_{r}^{2}$ in PSR J0348+0432 for $\chi = 1$ (Red), $\chi= 3$ (Blue), $\chi = 5$ (Green), $\chi=7$ (Purple) and $\chi=10 $ (Black) (considering $\alpha=0.5$)}
\end{figure}

\begin{figure}
\centering
\includegraphics[width=8 cm,height= 5 cm]{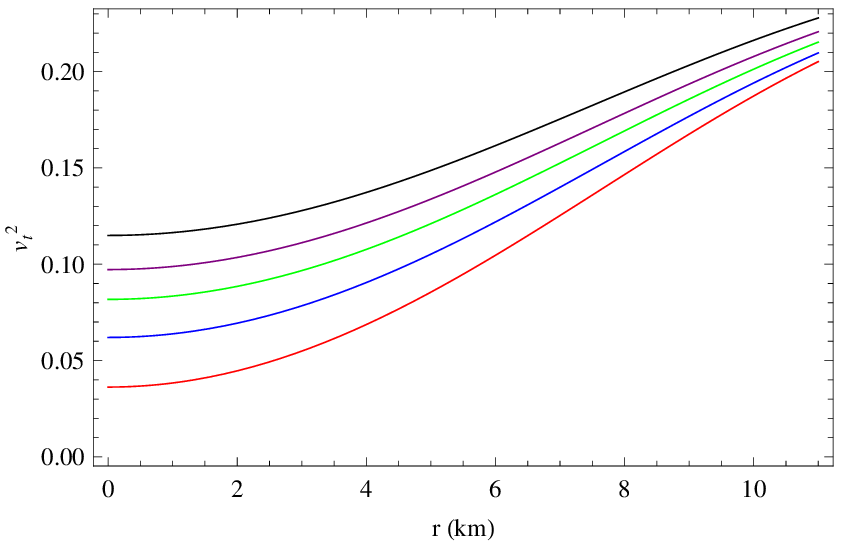}
\caption{Radial variation of $v_{t}^{2}$ in PSR J0348+0432 for $\chi = 1$ (Red), $\chi= 3$ (Blue), $\chi = 5$ (Green), $\chi=7$ (Purple) and $\chi=10 $ (Black) (considering $\alpha=0.5$)}
\end{figure}

\begin{figure}
\centering
\includegraphics[width=8 cm,height= 5cm]{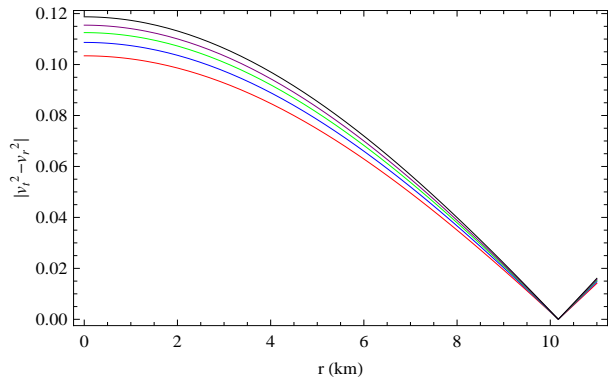}
\caption{Radial variation of $|v_{t}^{2}- v_{r}^{2}| $ in PSR J0348+0432 for $\chi = 1$ (Red), $\chi= 3$ (Blue), $\chi = 5$ (Green), $\chi=7$  (Purple) and $\chi=10 $ (Black) (considering $\alpha=0.5$)}
\end{figure}

\subsubsection{\bf Adiabatic index}

The stiffness of the EoS for given energy density is characterised by adiabatic index which has significant importance for understanding  the stability of relativistic as well as non-relativistic compact objects. Chandrasekhar   began the study of the dynamical stability against infinitesimal radial adiabatic perturbation of the stellar system. It is estimated that the magnitude of the adiabatic index should be greater than $\frac{4}{3}$ in the interior of a dynamically stable stellar object. For anisotropic fluid distribution the   adiabatic index is given by,
\begin{equation}
\Gamma= \frac{\rho+p_{r}}{p_{r}} \frac{dp_{r}}{d\rho}.
\end{equation}
The  radial variation of the adiabatic index is plotted  in Fig. (13) for different values of $\chi$. The stellar models obtained here are found to have dynamical stability as  $\Gamma \ge \frac{4}{3}$. The stellar models are stable against infinitesimal radial adiabatic perturbations.
In Fig. (14) we plot the radial variation of adiabatic index ($\Gamma$)  for different  values of the  parameter $ 0 < \alpha <0.5$  with $\chi=1$.  We note acceptable range $ 0 < \alpha < 0.3$ for $\chi= 0.5$ and $ 0 < \alpha <0.5$ for 
  $\chi= 1, \; 1.5. \, 2$.  Thus on increasing $\chi$ the acceptable  range of $\alpha$ remains same for anisotropic  star. Thus in $f(R,T)$ modified gravity we get an  upper bound on  $ \alpha$  for $\chi>0$ .

\subsection{\bf Energy conditions of the stellar model in the  $f(R,T)$ gravity}

The energy conditions play a crucial role in determining the observe normal or exotic nature of matter inside the stellar model. The
energy conditions are  null (NEC), dominant (DEC), strong (SEC) and weak energy conditions (WEC). in an 
 anisotropic fluid distribution are expressed as follows:
\begin{equation}
{\it NEC:}\;  \rho \ge 0,
\end{equation}
\begin{equation} 
{\it WEC1:}\;  \rho+p_{r} \ge 0,  \;\;\;\; {\it WEC2:} \rho+ p_{t} \ge 0,
\end{equation}
\begin{equation}
{\it SEC:}\;  \rho+p_{r}+2 p_{t} \ge 0,
\end{equation}  
\begin{equation}
{\it DEC1:}\;  \rho - p_{r} \ge 0, \;\;\;\;  {\it DEC2:}  \;\rho - p_{t} \ge 0,
\end{equation}
The evolution of all the energy conditions against the radial coordinate $r $ for the compact stellar structure is studied here for different $\chi$ with $\alpha=0.1$ in $f(R,T)$-gravity. These are shown graphically in the Figs. (15)- (19).

\begin{figure}
\centering
\includegraphics[width=8 cm,height= 5 cm]{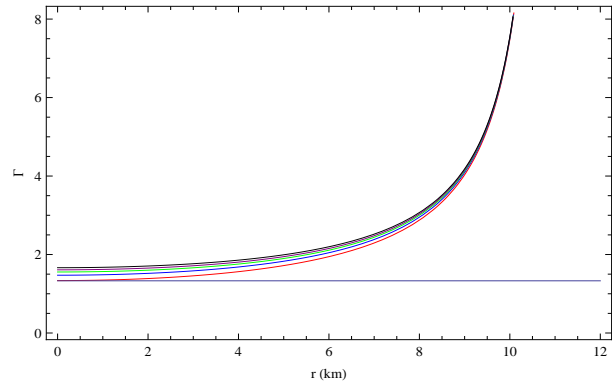}
\caption{Radial variation of $\Gamma$ in PSR J0348+0432 for $\chi = 1$ (Red), $\chi= 3$ (Blue), $\chi = 5$ (Green), $\chi=7$ (Purple) and $\chi=10 $ (Black) (considering $\alpha=0.5$)}
\end{figure}

\begin{figure}
\centering
\includegraphics[width=8cm,height= 5 cm]{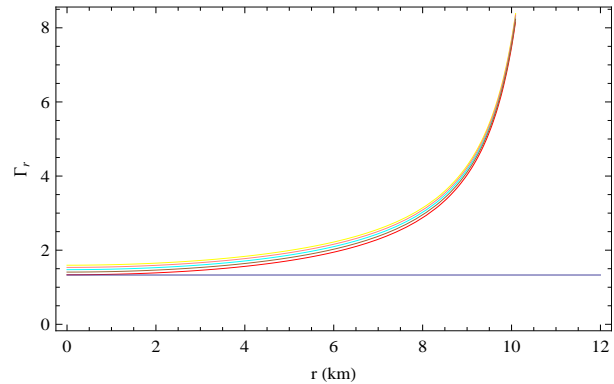}
\caption{Radial variation of $\Gamma$ in PSR J0348+0432 for $\alpha= 0.1$ (Yellow), $\alpha = 0.2 $ (Pink), $\alpha=0.3$(Cyan), $\alpha=0.4$(Brown) and $\alpha= 0.5 $ (Red) (considering $\chi=1$)}
\end{figure}

\subsection{\bf Stellar Mass - Radius Relation}
 For a static spherically symmetric stellar models with anisotropic fluid Buchdahl found a limit on the mass to radius ratio, $\it i.e.$ $\frac{2M}{R}< \frac{8}{9}$ \cite{37}. In this section we analyze  graphical behaviour of the mass- radius relation for different values of the parameters. The effective mass is given by
\begin{equation}
 m(r) = \int_{0}^{r} 4 \pi r'^{2}\rho dr'.
\end{equation}
We consider PSR J0348+0432  with observed mass equal to $M= 2.01$ $\pm$ 0.04 $M_{\odot}$. Now plotting the observed mass in
the mass-radius curve in Fig.(20), it is found that one can predict the variation of the size of a compact object for different $\chi$ values. In Fig.(20) it is shown that for a given mass of known object, the radius increases for the increasing values of  $\chi$, thus the compactness factor of the star decreases. Thus, we can state that for lower values of $\chi$ we can find more dense object comparatively. The mass function is regular at the center of the compact stellar structure. As it is not yet measured the radius of a star accurately, many aspects of a compact object may be understood once the mass and radius are determined accurately.

\subsection{\bf Class of Stellar Models with EoS}

The different physical parameters $a$, $ b$, $C$ of Finch-Skea metric given by eqs.  (\ref{31}) and (\ref{32}) are determined using  the boundary conditions, satisfying the criterion for a physically realistic stellar object. We tabulated different metric parameters  in Tables-1 and 2 for PSR J0348 + 0432  which admits different class of stellar model in $f(R,T)$ gravity.  The value of $C$ is calculated  for a particular stellar object which is  independent of $\chi$ and $\alpha$. Considering  $C =0.009664$, we determined the parameters  for PSR J0348 + 0432 whose observed mass $M= 2.01$ $\pm$ 0.04 $M_{\odot}$ and radius, $R= 11 km$. In Table-1,  values  of $a$ and $b$  for different $\chi$ at $\alpha=0.5$ and in Table- 2, the variations of $a$ and $b$  are displayed for different $\alpha$ taking $\chi=1$.  We also tabulated parameters for  different known sources namely, Vela X-1, 4U 1820-30, Cen X-3, LMC X-4, SMC X-1 with their precise estimated mass.  For  $\chi = 1$ and $\alpha=0.5$ we estimated the permissible radii.  It may be pointed out here that if the values of the parameters $\chi $ and $\alpha$ are taken different then for a given mass  one estimates the  radii which is different from the estimated value in the Table-2. As the radius of a star can not be measured precisely, we can predict the radius in the models. The  predicted radii  in the modified gravity with FS-geometry permits very compact objects namely, neutron stars, strange stars.

\subsection{\bf Equation of State}
The variation of the energy-density and radial pressure are plotted in Figs. (1) and (2) from which we determine functional form by best fitting the curve.
  Here we determine the best fit relation between $\rho$ and $p_{r}$, the expressions so obtained for different $\chi$ values have been listed in Table-1 for a given $\alpha$. In Fig.(21) we plot   $p_{r}$ vrs. $\rho$ for PSR J0348+0432 with different $\chi$ for a given $\alpha$. It is found that EoS for PSR J0348+0432 is non-linear for $\chi=1$  and $\chi=3$, but linearity develops in Fig. (21) as the values of $\chi$ increases. It is shown that quadratic fitting of the EoS curve is better than a linear one for lower value of $\chi$.  Thus the MIT Bag model representing the EoS in a compact star is not suitable in a compact object with FS geometry, it predicts a non-linear EOS. 
  
\begin{table}
\begin{center}
\begin{tabular}{|c|c|c|c|} \hline
$\chi$ & b & a & EoS   \\ \hline
1 & 0.21193& 0.31475 & $p_{r} = - 0.0000627+ 0.2174 \rho - 38.72 \rho^{2}$     \\  \hline    
3 & 0.22385& 0.26512 & $p_{r} = - 0.0000526+ 0.2139 \rho - 24.96 \rho^{2}$      \\  \hline
5 & 0.23221& 0.23033 & $p_{r} = - 0.00004291+ 0.20162 \rho$                       \\  \hline
7 & 0.23840& 0.20459 & $p_{r} = - 0.00004107+ 0.21631 \rho$                  \\  \hline
10& 0.24515& 0.17649 & $p_{r} = - 0.00003744+ 0.23221 \rho$                 \\  \hline
\end{tabular}
\caption{Physical parameters for PSR J0348+0432 in $f(R,T)$ gravity for $\alpha = 0.5$.}
\end{center}
\end{table}

\begin{table}
\begin{center}
\begin{tabular}{|c|c|c||c|c|c|} \hline
$\alpha$ & b & a  & $\alpha$ & b & a \\ \hline
0.1 & 0.26914 & 0.28295 & - 0.1 & 0.28035 & 0.30121  \\  \hline
0.2 & 0.25967 & 0.27915 & - 0.3 & 0.28273 & 0.32754   \\  \hline
0.3 & 0.24723& 0.28104 & - 0.5 & 0.27767 & 0.35765    \\  \hline
0.4 & 0.23148 & 0.29136 & - 0.7 & 0.26633 & 0.38909   \\  \hline
0.5 & 0.21193 & 0.31475  & -1.0 & 0.23950 & 0.43558    \\  \hline
\end{tabular} 
\caption{Anisotropy and Metric parameters  for PSR J0348+0432 in modified gravity for  $\chi=1$.}
\end{center}
\end{table}

\begin{table}
\begin{center}
\tabcolsep=0.11cm
\begin{tabular}{|c| c| c| c| c| c|} \hline
Stars & Mass ($M_{\odot}$) & $b$ & $a$ & $C$ &  b (km.) \\ \hline
Vela X-1 & 1.77 $\pm$ 0.08 & 0.24665& 0.31616 & 0.00783 & 10.88 \\ \hline
4U 1820-30 & 1.58 $\pm$ 0.06 & 0.26769& 0.31601 & 0.00716 & 10.52 \\ \hline
Cen X-3 & 1.49$\pm$ 0.08 & 0.27895& 0.31616 & 0.00783 & 10.88 \\ \hline
LMC X-4& 1.29 $\pm$ 0.05 & 0.29876& 0.31439 & 0.00631 & 9.926\\ \hline
SMC X-1 & 1.04$\pm$ 0.09 & 0.3250& 0.31172 & 0.00569 & 9.301\\ \hline
\end{tabular}
\caption{Numerical values of physical parameters for different compact object for $\alpha=0.5$ and $\chi=1$ where $b$ represents the predicted radius of the pulsars.}
\end{center}
\end{table}
\begin{figure}
\centering
\begin{minipage}[t]{5 cm}
\centering
\includegraphics[width= 7cm,height= 4 cm]{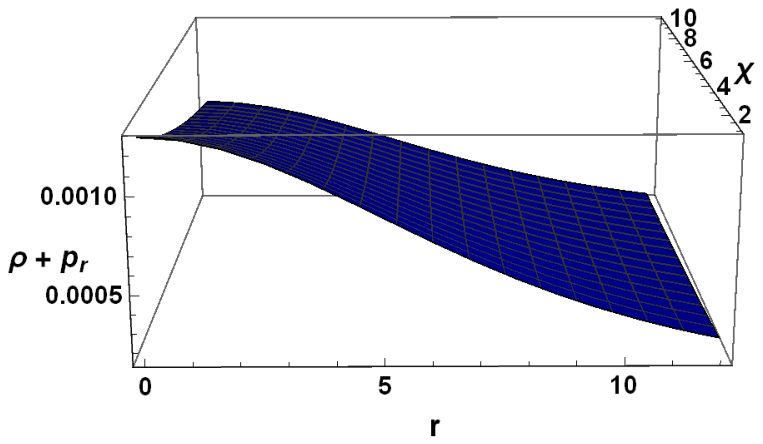}
\caption{WEC 1}
\end{minipage}
\hspace{3 cm}
\begin{minipage}[t]{5 cm}
\centering
\includegraphics[width= 7cm,height= 4cm]{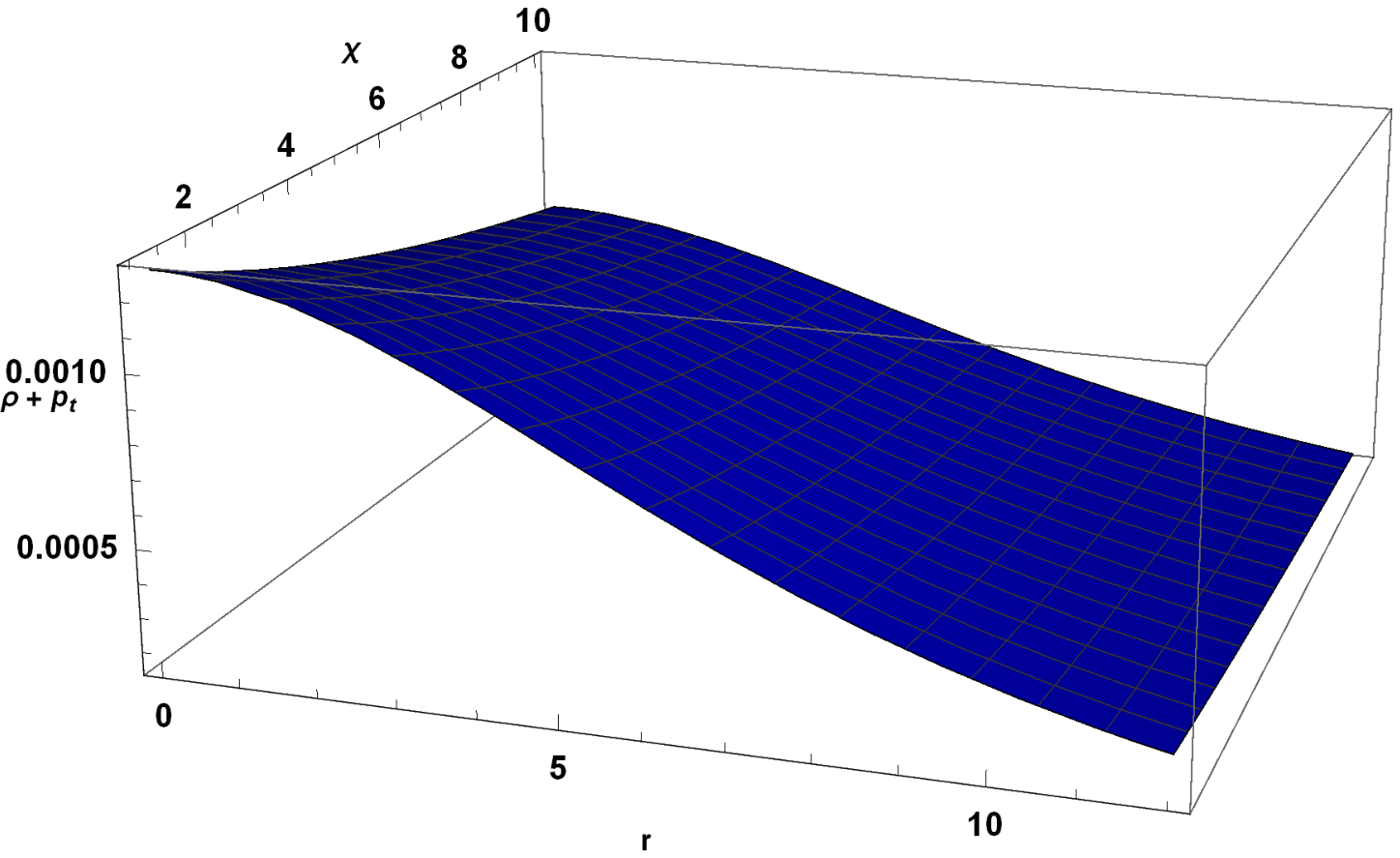}
\caption{WEC 2}
\end{minipage}

\begin{minipage}[t]{5 cm}
\centering
\includegraphics[width= 7cm,height= 4cm]{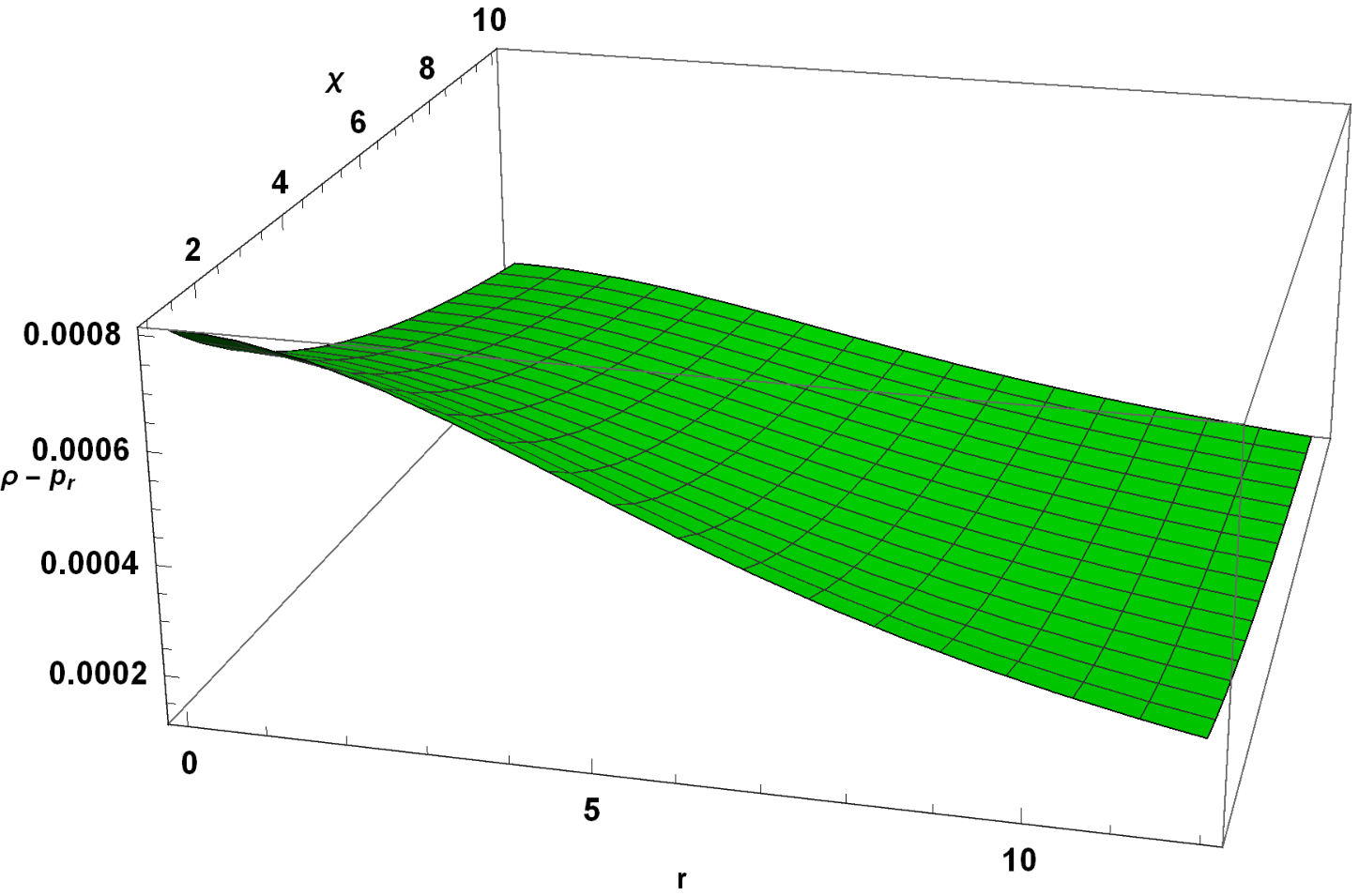}
\caption{DEC 1}
\end{minipage}
\hspace{3 cm}
\begin{minipage}[t]{5 cm}
\centering
\includegraphics[width= 7cm,height= 4cm]{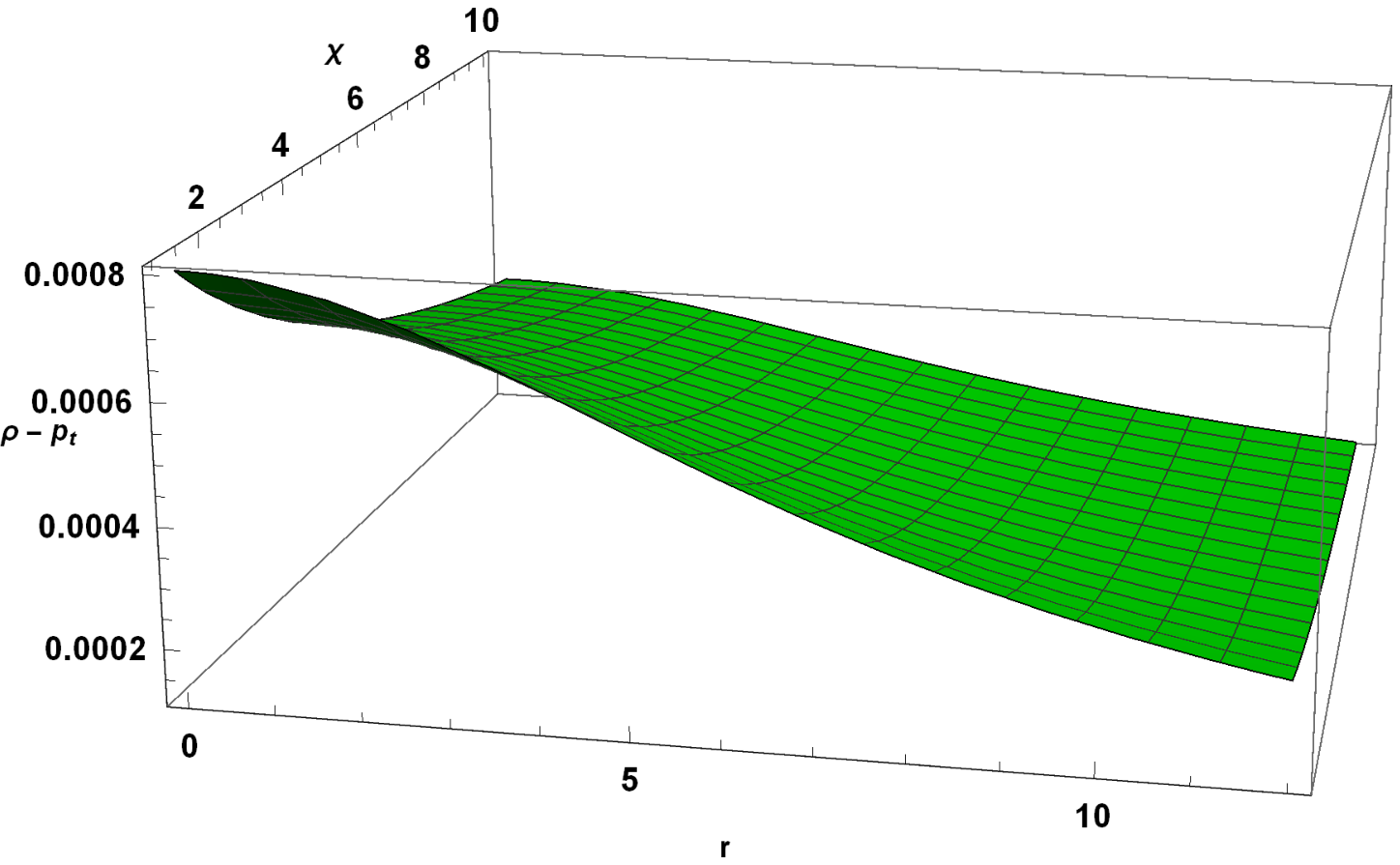}
\caption{DEC 2}
\end{minipage}

\begin{minipage}[t]{5 cm}
\centering
\includegraphics[width= 7cm,height= 4cm]{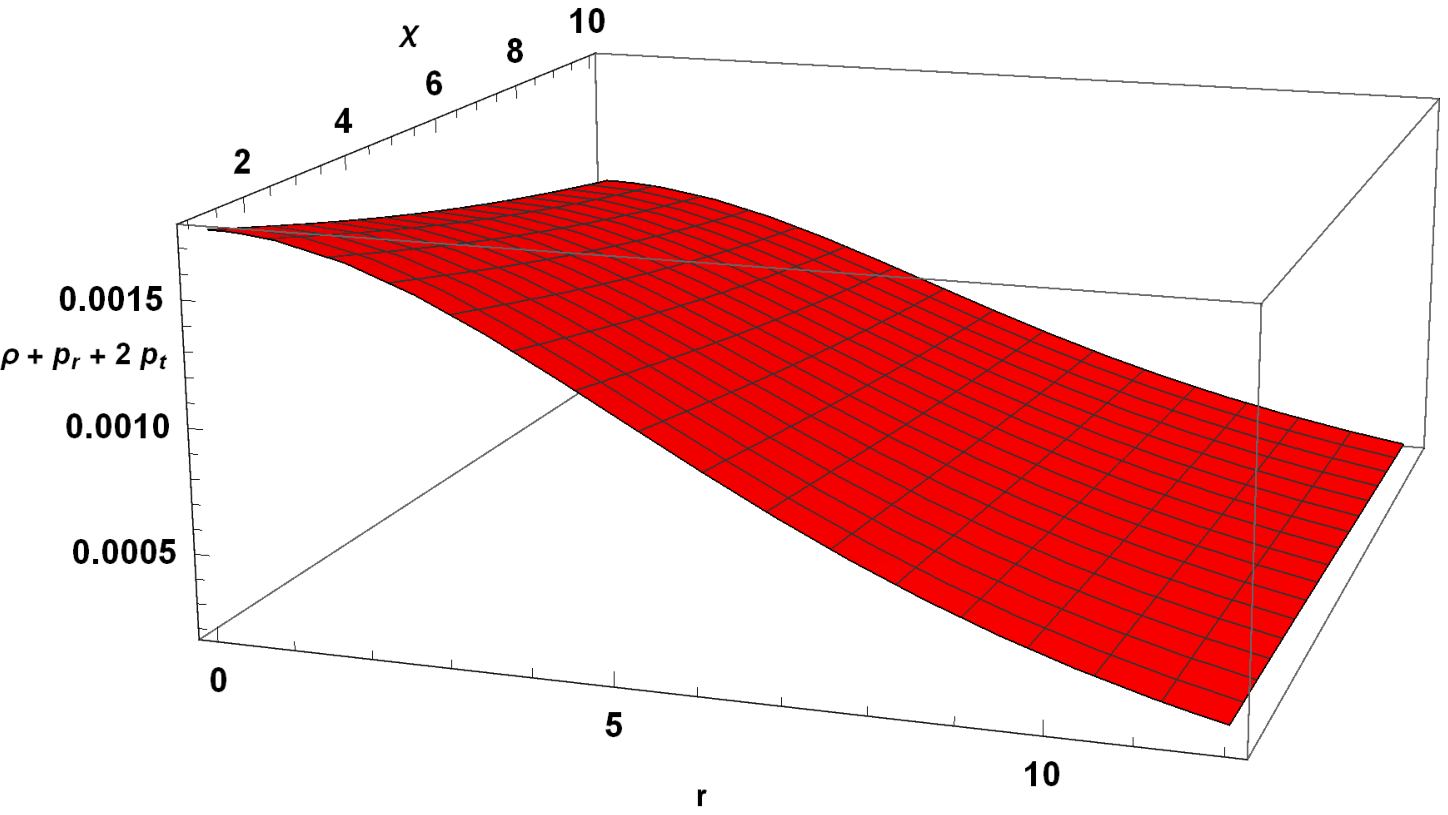}
\caption{SEC}
\end{minipage}

\end{figure}

\begin{figure}
\centering
\includegraphics[width=7cm,height= 5cm]{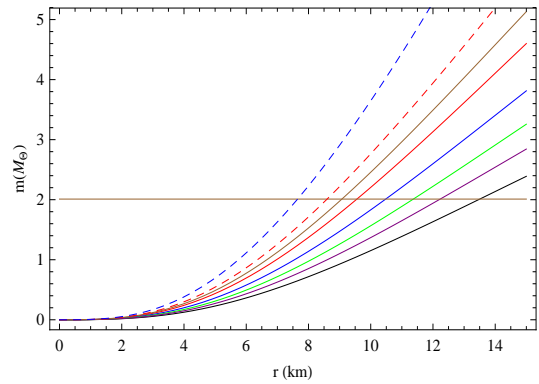}
\caption{Mass - Radius relation in PSR J0348+0432 for $\chi = -3$ (Blue Dashed) and $\chi = - 1$ (Red Dashed)taking $\alpha= - 0.5$, $\chi = 0 $ (Brown) $\chi = 1$ (Red), $\chi= 3$ (Blue), $\chi = 5$ (Green), $\chi=7$ (Purple) and $\chi=10 $ (Black) for $\alpha=0.5$}
\end{figure}

\begin{figure}
\centering 
\includegraphics[width=8 cm,height= 5cm]{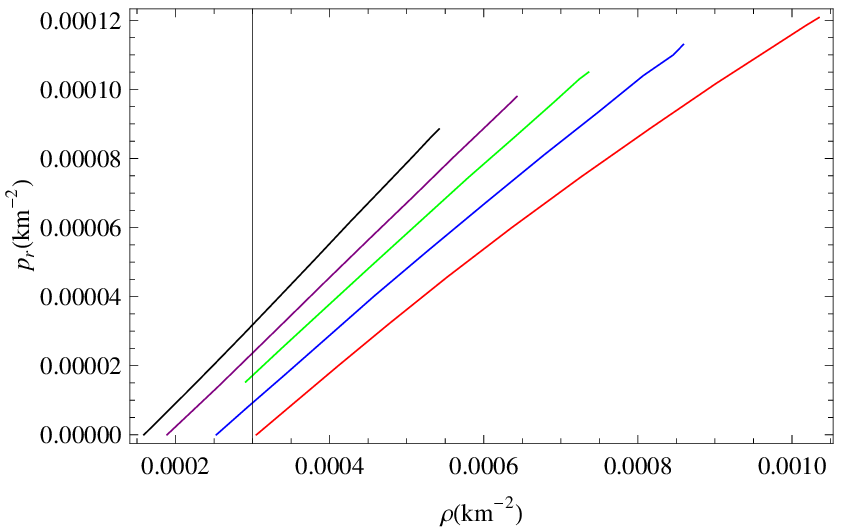}
\caption{EoS in PSR J0348+0432 for $\chi = 1$ (Red), $\chi= 3$ (Blue), $\chi = 5$ (Green), $\chi=7$ (Purple) and $\chi=10 $ (Black)  for $\alpha=0.5$}
\end{figure}

\section{Discussion} 
In the paper we obtain a class of relativistic solutions for  compact objects   in hydrostatic equilibrium in a modified gravity    $f(R,T)=R+2\chi T$.  Since the field equations are highly complex we adopt a technique to project the field equation in a second order differential equation. The anisotropic stellar models are constructed here. We analyze the stellar models numerically and predicted the EoS of matter inside the compact objects assuming a modified   Finch-Skea metric.  For $\chi=0$ it corresponds to GR  and it represents isotropic stellar configuration \cite{26}.
In the  modified gravity it is found that stellar models represent anisotropic uncharged compact objects always unless $\chi = -4 \pi$. It is also found that realistic stellar models are permitted  for a given range of values of anisotropy  $-1 < \alpha < +0.5 $ which are   stable. We study the physical features of the compact stars of known observed mass.
 As the EoS in a compact object  at extreme condition of density is not yet known we adopted a different technique, considering a modified Finch-Skea ansatz in the framework of modified gravity the EoS is predicted with different values of the model parameters. For a known stars  its mass is precisely known but not its radius, consequently the parameters of the model are determined analytically to find the probable radius  and the EoS of the matter inside the compact objects. A precise measurement of radius will be helpful for 
 We note the following :\\

$(i)$  The radial variation of energy density, radial pressure and tangential pressure plotted in Figs. (1) - (4) show that  they are maximum at the origin which however decrease away from the centre.   The radius of the star is determined from  the condition that the radial pressure vanishes at the surface. The coupling  parameter  $\chi$ in the gravitational action is playing an important  role to accommodate anisotropic compact objects. We note that  both the central density and  pressure decreases as  $\chi$ is increased.  We note that as $\chi$ is decreased it accommodates a more dense star.\\

$(ii)$ There is no physical and mathematical singularities as  the radial variation of the radial pressure ($p_{r}$) and tangential pressure ($p_{t}$) shown in Fig. (2) - (5) are positive and regular at the origin. \\

$(iii)$ An isotropic stellar configuration is obtained for  $\alpha=0$  in eq. (31).  For PSR J0348+0432, the radial variation of  $\Delta$ for different  $\chi$  in Fig.(8) shows that $\Delta >0$ {\it i.e.}  $p_{t}> p_{r}$  for $\chi> 0$ which  implies that the anisotropic stress is directed outwards. There exists a repulsive gravitational force that allows the formation of super massive star in this case.  \\

$(iv)$  For $\chi=1$ the plot of radial variation of $\Delta$ for different values of $\alpha$   in Fig. (9) shows that it admits stellar models with   $p_{r}> p_{t}$ indicating the formation of ultra compact objects. It is also noted that 
the range $ 0.5< \alpha< 1$  is not suitable as no stable configuration allowed. In $f(R,T)$  gravity,  the anisotropy is small near the centre which however attains  maximum value at the surface. 
For  PSR J0348 + 0432 we determine the values of $chi$  for which $\Delta < 0 $. It is found that   (i) $\alpha= - 1.0$ in the range  $-1.4 <\chi <  - 1$, (ii)  $\alpha= - 0.5$   in the range  $-3.5 <  \chi < -1$ , (iii)  $\alpha= - 0.2$, in the range $-4.4 < \chi < -1$, thus as $\alpha$ is decreased the lower  value of  $\chi$ is increased. \\

$(v)$ The radial variation of the adiabatic index $\Gamma$ plotted in Fig.(13) shows that  the stellar models are stable as it satisfies the  Buchdhal limit $\chi > \frac{4}{3}$. A class of relativistic solutions are obtained here for anisotropy lying in the range   $0< \alpha < 0.5$ which permits stable stellar models evident from the Fig.(14).   The $f(R,T)$-gravity with modified FS-metric ansatz permits  anisotropic star in four dimensions, it is different from that of GTR result where it accommodates stars with isotropic pressure. We obtain  upper bounds on  anisotropy  $\alpha$  for different $\chi$ for anisotropic stars. \\

$(vi)$ All the energy conditions, $viz.$, (a) Null energy condition (NEC), (b) Weak energy condition (WEC) and (c) Strong energy condition (SEC) drawn in Figs. (15) - (19)  are satisfied. Thus no exotic matter required for building stellar models. \\

$(vii)$The mass-radius relation of PSR J0348+0432 plotted in Fig.(20) for different values of $\chi$ shows that for a given mass of the compact object, the radius increases for an increasing value of  $\chi$.  Thus for lower $\chi$, the models accommodates very compact object as the compactness factor $\left(\frac{M}{b}\right)$ increases (where $b$ is the radius of a star). \\

$(viii)$ We constructed stellar models for PSR J0348 + 0432  without pre-assuming EoS.  Instead we assume modified FS metric ansatz to determine EOS with  the metric coefficients $a$, $b$ and gravitational coupling parameter $\chi$ for an anisotropic configuration with $\alpha=0.5$.  The probable EoS are tabulated in  Table-1, it is evident that both linear and quadratic EoS are obtained. The numerical  fitting of the pressure and density curves show
that the goodness of fit for the quadratic fitting  is better than that of the linear one for lower values of $\chi$.  
However, the linear EoS obtained here are different from that corresponds to MIT bag model \cite {23,24}.  
 The EoS for matter interior to a compact star in modified gravity is predicted here which is non-linear for massive star.
 In Table- 2, we displayed  $a$ and $b$  for different anisotropy ($\alpha$) with $ \chi=1$ for a stable stellar configuration. The anisotropy lies in the range $ -1.0 < \alpha <  0.5$  for $\chi= 1$ in a stable stellar model. \\

$(ix)$ For  observed masses of the  pulsars namely, {\it Vela X-1, 4U 1820-30, Cen X-3, LMC X-4, SMC X-1,}  we determine the values of $a$, $b$ $C$ with $\chi=1$ stable anisotropic models are shown for  anisotropy $\alpha = 0.5$. A class of relativistic solutions are obtained for different anisotropic pressure inside the star. The predicted radius of the above pulsars for the parameters are displayed in Table-3. However, varying the values of $a$, $b$ $C$,  it is possible to obtain stable anisotropic stellar models with different coupling parameter and anisotropy. It is possible to estimate the corresponding radius which lies in the range $(10  \sim 14)$ $km.$ for a stable neutron star. \\
 Thus a class of new relativistic solutions are found in $f(R, T)$ gravity with FS ansatz which are useful for building stellar models. The precise measurement of radius of a neutron star in future will be useful to accept the modification incorporated in the gravitational action for building stellar models which can dig out information on the matter inside the star at extreme terrestrial condition. It is found that the EoS for a compact object with modified FS ansatz
 is non-linear.

\section{Acknowledgements}

SD is thankful to UGC, New Delhi for financial support.  AC would like to thank University of North Bengal for awarding Senior Research Fellowship. The authors would like to thank IUCAA Resource Center, NBU for extending research facilities. BCP would like to thank DST-SERB Govt. of India (File No.: EMR/2016/005734) for a project.

\end{document}